\documentclass[smallextended]{svjour3}       
\smartqed  
\usepackage{graphicx}
\usepackage{euscript,amsmath,amssymb,amsfonts,graphicx,fullpage,bm,color}

\usepackage{natbib}

\numberwithin{equation}{section}

\usepackage{epstopdf}

\newcommand{\e}{{\rm e}}
\renewcommand{\d}{{\rm d}}

\newcommand{\D}{\displaystyle}
\newcommand{\mc}{\mathcal }
\newcommand{\ve}{\varepsilon}

\newcommand{\A}{\mathcal{A}}
\newcommand{\B}{\mathcal{B}}
\newcommand{\M}{\mathcal{M}}
\newcommand{\N}{\mathcal{N}}
\newcommand{\U}{\mathcal{U}}

\newcommand{\wee}{\bar{w}_{ee}}
\newcommand{\wei}{\bar{w}_{ei}}
\newcommand{\wie}{\bar{w}_{ie}}

\begin{document}

\title{Encoding certainty in bump attractors}

\titlerunning{Certainty in bumps}        

\author{Sam Carroll \and Kre\v{s}imir Josi\'{c} \and Zachary P. Kilpatrick}

\institute{S.R. Carroll \at
              Department of Mathematics, University of Houston, \\
              Houston TX 77204 USA \\
              \email{srcarroll314@gmail.com} 
              \and
              K. Josi\'{c} \at
              Department of Mathematics, University of Houston, \\
              Houston TX 77204 USA \\
              \email{josic@math.uh.edu}
              \and
              Z.P. Kilpatrick \at
              Department of Mathematics, University of Houston, \\
              Houston TX 77204 USA \\
              \email{zpkilpat@math.uh.edu}    
}

\date{Received: date / Accepted: date}

\maketitle

\begin{abstract}
Persistent activity in neuronal populations has been shown to represent the spatial position of remembered stimuli. Networks that support bump attractors are often used to model such persistent activity.   Such models usually exhibit translational symmetry.  Thus activity bumps are neutrally stable,  and perturbations in position do not decay away. We extend previous work on bump attractors by constructing model networks capable of encoding the certainty or salience of a stimulus stored in memory.  Such networks support bumps that are not only neutrally stable to perturbations in position, but also  perturbations in amplitude. Possible bump solutions then lie on a two-dimensional attractor, determined by a continuum of positions and amplitudes. Such an attractor requires precisely balancing the strength of recurrent synaptic connections. The amplitude of activity bumps represents certainty, and is determined by the initial input to the system.  Moreover, bumps with larger amplitudes are more robust to noise, and over time provide a more faithful representation of the stored stimulus. In networks with separate excitatory and inhibitory populations, generating bumps with a continuum of possible amplitudes, requires tuning the strength of inhibition to precisely cancel background excitation.

\end{abstract}

\section{Introduction}

Neuronal populations in many cortical areas exhibit sustained activity during the delay period in a spatial working memory task \citep{wang01,curtis06}. Groups of cells responsive to the presence of a stimulus that needs to be stored in memory can remain active after the stimulus is removed \citep{rao00,vijayraghavan07}. Which subset of neurons is active depends on the spatial location of the cue~\citep{funahashi89}. Such sustained activity has been observed in prefrontal cortex \citep{goldmanrakic95}, parietal cortex \citep{pesaran02}, as well as superior colliculus \citep{basso97}. 

Such persistent elevation in firing rates is captured in model networks by ``bumps" of activity.  The peaks of these activity bumps represent the remembered location of the cue \citep{compte00,durstewitz00,gutkin01}. Maintaining a stable activity bump during the delay is hence crucial for representing the remembered cue \citep{brody03}. The recurrent architecture of the local neuronal networks appears to play a crucial role in maintaining such selective activation \citep{constantinidis04}. Tuned excitatory neurons reciprocally connect to one another with both fast and slow synapses \citep{wang01}. In addition, inhibitory cells broadly project back to the rest of the network keeping spatial tuning sharp \citep{rao00}.   Understanding how synaptic architecture can be tuned to produce reliable bumps is essential for understanding the mechanism behind  spatial working memory. 

Models that can store a continuous range of spatial locations typically possess solutions that are neutrally stable\citep{amari77,seung96,brody03}. Due to the neutral stability of such attractors,  perturbations that change the location of a bump of activity do not decay away~\citep{amari77,camperi98,compte00}. Aside from experimentally-introduced distractors in spatial working memory experiments \citep{miller96}, cue memories can also be degraded by internal variability within cortical networks~\citep{faisal08}. Stochastic models show that such variability causes bump attractors to wander diffusively, due to their inherent neutral stability \citep{camperi98,compte00,laing01,kilpatrick13}. Psychophysical studies show that errors made recalling remembered spatial locations scale roughly linearly with delay time, suggesting the remembered location may diffuse in time \citep{white94,ploner98}. Also,  heterogeneities in the spatial structure of the underlying neuronal network can further degrade the relation between the stored memory and the initial cue \citep{seung96,renart03,itskov11,hansel13}. One solution to this problem is to structure the spatial 
arrangement of excitatory synapses \citep{kilpatrick13,kilpatrick13b} to make networks robust to dynamic and static parametric perturbations. Thus, the spatial organization of synaptic architecture can play a major role in accurately 
encoding stimuli for future recall.

To explore the relation between network architecture and the neural computation underlying working memory, we consider bump attractor networks capable of encoding cue {\em certainty}. We define {\em certainty} as the likelihood that the presented cue was faithfully communicated to the network generating delay period activity.  A number of experiments  have shown that the certainty of a decision can be encoded by the instantaneous firing rates of neurons in medial temporal cortex~\citep{shadlen98,gold02,beck08,kiani09}. In this way, the activity of a network can represent the encoded signal as well as the likelihood that the encoded signal accurately represents reality \citep{zemel98}. We introduce this notion here in the context of spatial working memory. Recordings from superior colliculus by \cite{basso97} suggest that increased uncertainty in a remembered cue position is represented by lower neural activity during the delay period. Conversely, work by \cite{meyer11} shows that training in a spatial working memory task that leads to improved performance is also accompanied by a rise in delay period firing rates. These observations suggest that  certainty about stimulus location in a spatial working memory task may be represented by the level of neural activity during the delay period.

Like cue position, the degree of certainty in a signal is an analog quantity. Phenomenological models of decision making in the presence of two alternatives also frequently exhibit line attractors.  In such models the state along the line attractor represents the likelihood, or certainty, that one of the choices is correct~\citep{bogacz06}. This can be accomplished in more biophysically realistic models by choosing synaptic time constants to match the slope of the input-output relationship of a firing rate model \citep{wang02}. Precisely balancing the rate of feedback excitation with the timescale of synaptic decay leads to a model that behaves as a pure integrator. In the absence of external inputs, networks that behave as pure integrators can store the value of a continuous variable \citep{goldman03}. Similar tuning can also be accomplished in mutually inhibitory rate models for parametric working memory \citep{machens05,polk12}.

We build on these ideas to study spatial working memory networks that can encode certainty. Typically, bump attractor networks only possess a single stable bump amplitude at each orientation \citep{amari77,ermentrout98}. We explore networks that support a continuum of  bump amplitudes at each orientation. These are dynamical systems that contain two-dimensional attractor surfaces: One dimension corresponded to the amplitude, and the other the position of the bump. The system therefore exhibits a plane attractor. To accomplish this, excitation and inhibition must be balanced, and the shape of the synaptic input to output firing rate function chosen appropriately \citep{amari77}. Namely, there is a monotonic relationship between the total synaptic excitation and total synaptic inhibition that must be maintained to represent a continuum of possible bump amplitudes.

Including a certainty code in networks for bump attractors has several consequences. First, the strength of the original input can be encoded in the amplitude of the bump. Bumps with larger amplitudes stay closer to their original position when the effects of noise are considered during the storage period. However, since certainty is encoded in the amplitude of the bump, memory of the original certainty can also be degraded by dynamic noise. Also, arbitrarily weak inputs can still be stored as a bump attractor, which is not the case in networks that support a single stable bump amplitude \citep{amari77,ermentrout98}. Using a stochastic neural field model of a bump attractor network with noise, we can develop explicit formulae for most of these results using asymptotic methods.

\section{Bump attractor networks}
\label{model}

Bump attractor networks were originally developed as general models of recurrent neuronal circuits that can support spatiotemporal patterns of  activity \citep{wilson73,amari77}. Since then they have been used to represent activity subserving spatial working memory \citep{camperi98,compte00} and visual orientation processing \citep{benyishai95}. We consider a spatially organized neural field model where the positions of neurons correspond to their preferred stimulus orientation. We focus on a ring architecture, but we believe these ideas will extend to more general models. 

Consider a single population which incorporates local excitation and broadly tuned inhibition \citep{amari77,benyishai95,ermentrout98}
\begin{equation}\label{eq:singlenet}
\frac{\partial u(x,t)}{\partial t} = -u(x,t) + \int_{-\pi}^{\pi}w(x,y)f(u(y,t))dy+I(x,t).
\end{equation} 
Here $u(x,t)$ is the total synaptic input to spatial location $x \in [-\pi, \pi]$ at time $t$.  The integral kernel, $w(x,y)$, represents the synaptic feedback from the whole network which encodes the strength of connections from $y$ to $x$.  For a translationally symmetric synaptic weight function, $w(x,y) = \bar{w}(x-y)$, it can be shown that bump solutions will be neutrally stable to translations in position \citep{amari77,ermentrout98,veltz10}. This relies on the assumption of a spatial homogeneity in the net excitability of any particular neuron in the network \citep{renart03,kilpatrick13}.   For simplicity, we use a unimodal synaptic weight function
\begin{equation}\label{eq:singleweight}
w(x,y) = w_0 + w_1\cos{(x-y)},
\end{equation}
where $w_0$ represents the amplitude of broadly tuned inhibition, and $w_1$ represents the amplitude of locally tuned excitation. Results on the existence and stability of stationary bump solutions can easily be extended to synaptic weights with many more modes \citep{veltz10}. However, the anatomical structure of recurrent connectivity is not known at such a fine level of detail \citep{rao00}. We therefore use canonical functions that represent the short range excitation and broad inhibition known to exist.

The nonlinearity, $f,$ is the firing rate function which maps the synaptic inputs, $u,$ to a resulting fraction of active neurons
(or probability of activation of a single neuron). Typically, $f$ is a saturating, non-negative function ~\citep{coombes04,bressloff12}.  In this study, we consider a piecewise linear firing rate function of the form \citep{hansel98,pinto01b,kilpatrick10}
\begin{equation}\label{eq:firingrate}
f(u) = 
\begin{cases}
0, & \text{\indent if $u< \theta$} \\
s(u-\theta), & \text{\indent if $\theta \le u \le \frac{1}{s} + \theta$} \\
1, & \text{\indent if $u > \frac{1}{s} + \theta$},
\end{cases}
\end{equation}
where $s$ is the gain parameter, and $\theta$ the threshold. This choice will allow for a straightforward construction of a network capable of storing a continuum of bump amplitudes. In typical analyses of the spatiotemporal dynamics of neural fields, the parameters in Eq.~(\ref{eq:firingrate}) are chosen so that the underlying space-clamped system is bistable \citep{hansel98,pinto01b}. Since we assume the population $u$ is quite large, we make the assumption $\theta = 0$ throughout this study, so arbitrarily weak inputs always activate a small fraction of the population \citep{hansel98}.

In addition, we will analyze a two population network containing separate excitatory and inhibitory populations. The model we employ ignores inhibitory-inhibitory interactions -- the small proportion of inhibitory-inhibitory synaptic connections observed in prefrontal cortex is not expected to alter our results substantially \citep{somogyi98}. Thus, we consider the system of integro-differential equations \citep{pinto01}
\begin{eqnarray}\label{eq:einet}
\frac{\partial u(x,t)}{\partial t} & = & -u(x,t) + \int_{-\pi}^{\pi}w_{ee}(x,y)f(u(y,t))dy - \int_{-\pi}^{\pi}w_{ie}(x,y)v(y,t)dy + I(x,t) \nonumber \\
\tau\frac{\partial v(x,t)}{\partial t} & = & -v(x,t) + \int_{-\pi}^{\pi}w_{ei}(x,y)f(u(y,t))dy,
\end{eqnarray}
where $u(x,t)$ is the total synaptic input to the excitatory network and $v(x,t)$ is the total synaptic input to the inhibitory network.  The integral kernel, $w_{ee}$, is the synaptic strength of the excitatory network onto itself, $w_{ei}$ is the strength of the excitatory network onto the inhibitory network, and $w_{ie}$ is the strength of the inhibitory network onto the excitatory network.  Additionally, $\tau$ is the inhibitory time constant which denotes the speed at which inhibition acts on the excitatory population.  We will consider weight functions of the form
\begin{eqnarray}\label{eq:multiweight}
w_{ee}(x) & = & \wee(1+\cos{x}) \nonumber \\
w_{ei}(x) & = & \wei(1+\cos{x}) \nonumber \\
w_{ie}(x) & = & \wie
\end{eqnarray}
where $\wee$,$\wei$,$\wie > 0$.  These functions are non-negative. Constant inhibition was chosen both to represent
the broader tuning of inhibition compared to excitation, and to ease mathematical analysis. 

We note that in the limit of fast inhibition, $\tau \to 0$, Eq.~(\ref{eq:einet}) reduces to  
\begin{eqnarray}
\frac{\partial u(x,t)}{\partial t} & = & -u(x,t) + \left(w_{ee}(x) - w_{ie}(x)*w_{ei}(x)\right)*f(u(x,t)) + I(x,t) \nonumber \\
v(x,t) & = & \int_{-\pi}^{\pi}w_{ei}(x,y)f(u(y,t))dy.
\end{eqnarray}
Here the first equation is equivalent to the single population case in Eq.~(\ref{eq:singlenet}), and the second equation has no impact on stability.  Therefore,  the study of a single population can provide insight into the behavior of the  two population network.  

We first describe the general procedure for constructing stationary bump solutions with arbitrary firing rate and weight functions in the network (\ref{eq:singlenet}).  In the absence of external input, we look for the stationary bump solutions, $u(x,t)=U(x)$, by plugging into Eq.~(\ref{eq:singlenet}) and obtaining
\begin{equation}
U(x) = \int_{-\pi}^{\pi}w(x,y)f(U(y))dy.
\end{equation}
Since $U(x)$ must be periodic we expand it in a Fourier series
\begin{equation}\label{eq:U(x)}
U(x) = \sum_{k=0}^NA_k\cos{(kx)}+\sum_{l=1}^NB_k\sin{(kx)}
\end{equation}
where $N$ is the maximal mode. The assumption of there being a finite number of terms in the Fourier expansion for $U(x)$ relies on the weight function $w(x,y)$ having a finite Fourier expansion. This is reasonable since most typical smooth weight functions can be well approximated by a few terms in a Fourier series \citep{veltz10}. Doing so, allows us to always construct solvable systems for the coefficients of the bump and its stability. In the most general case for a spatially homogeneous weight kernel we write
\begin{equation}\label{eq:expandweight}
w(x,y) = \bar{w}(x-y) = \sum_{k=0}^Nw_k\cos{(k(x-y))} = \sum_{k=0}^Nw_k\left[\cos{(kx)}\cos{(ky)}+\sin{(kx)}\sin{(ky)}\right],
\end{equation}
so that
\begin{equation}
A_k = w_k\int_{-\pi}^{\pi}\cos{(kx)}f(U(x))dx, \text{\indent} B_l=w_l\int_{-\pi}^{\pi}\sin{(lx)}f(U(x))dx.
\end{equation}
Since the system is translationally symmetric, solutions centered at any position imply a translated solution of that same shape exists. In addition, the system (\ref{eq:singlenet}) will be reflection symmetric as well \citep{amari77}. With this in mind, we look solely for even solutions \citep{amari77,veltz10}, so that $B_l = 0$ for all $l$, and (\ref{eq:U(x)}) becomes
\begin{equation}
U(x) = \sum_{k=0}^NA_k\cos{(kx)}\label{eq:U(x)even}.
\end{equation}
By requiring self-consistency we have that
\begin{equation}
A_k = w_k\int_{-\pi}^{\pi}\cos{(kx)}f\left(\sum_{k=0}^NA_k\cos{(kx)}\right)dx.
\end{equation}

In general, numerical methods must be used to solve for the coefficients.  However for particular functions the solutions can be found analytically as we show in subsequent sections. In addition, we can typically compute the spectrum of the linear system governing perturbations of the bump \citep{coombes04,folias04,veltz10}, which yields relationships between the eigenvalues determining bump stability and parameters of $w$ and $f$. It is then straightforward to tune parameters to attain neutral stability along the two eigendirections of interest: one corresponding to translations of the bump (position) and the other corresponding to expansions/contractions of the bump (amplitude). We start by demonstrating this in a single population model.

\section{Neutral stability in a single population}
\label{single}

We first derive conditions for a network that supports bump solutions with a continuum of possible amplitudes.  We find that the parameters of the network must be tuned precisely. We find that recurrent excitation must be inversely proportional to the gain of the firing rate function.  The resulting network supports a continuum of bump amplitudes, and is capable of encoding certainty in the level of initial activation.  This initial activation is  controlled by both the duration of cue exposure, and the contrast (intensity) of the cue.  That is, longer cue times and/or higher contrast cue lead to higher initial amplitude, corresponding to higher certainty.  Finally, we examine the dynamics of bumps during the delay period using a stochastic neural field equation with additive white noise. We find that the spatial diffusion of the bump depends on its amplitude, and stronger initial activation results in more stable bumps.  This has interesting implications for working memory:  The greater the initial activation, the greater the certainty, and the better the stimulus is remembered during the delay period.

\subsection{Stationary Bump Solution}\label{stationary_single}
First we construct stationary bump solutions in the absence of external input and noise by plugging the stationary solution $u(x,t)=U(x)$ into Eq.~(\ref{eq:singlenet}) assuming a unimodal weight function given in Eq.~(\ref{eq:singleweight}).  We obtain
\begin{eqnarray}\label{eq:Ufixed_single}
U(x) = \int_{-\pi}^{\pi}\left(w_0+w_1\cos(x-y)\right)f(U(y))dy.
\end{eqnarray}
We then write, 
\begin{equation}
U(x) = A_0 + \sum_{k=1}^NA_k\cos{(xk)} + \sum_{k=1}^NB_k\sin{(xk)},
\end{equation}
where $N$ is the maximum mode of $U(x),$ and find that
\begin{eqnarray} \label{eq:singlecoeff}
A_0 & = & w_0\int_{-\pi}^{\pi}f(U(y))dy, \nonumber \\
A_1 & = & w_1\int_{-\pi}^{\pi}\cos{y}f(U(y))dy, \nonumber \\
B_1 & = & w_1\int_{-\pi}^{\pi}\sin{y}f(U(y))dy,
\end{eqnarray}
and $A_k = B_k = 0$ for $k \ne 0,1$.  In particular we would like to restrict our analysis to even stationary bump solutions, and we thus set $B_1 = 0$. Since $f(U(x))$ is then even, the last integral in (\ref{eq:singlecoeff}) will be zero, making the equation for $B_1$ self-consistent. 

Given the piecewise linear firing rate function in Eq.~(\ref{eq:firingrate}), it is useful to set certain conditions on the stationary bump solution $U(x)$. To start, we consider bump solutions where $0 \le U(x) \le \frac{1}{s}$. In other words, we will consider strictly positive bumps $U(x)$ whose peaks lie below the saturating threshold of (\ref{eq:firingrate}). We will now show that excitation must properly balance inhibition as well as the gain of the firing rate function in order to attain a line of neutrally stable bump amplitudes. When $0 \leq U(x) \leq \frac{1}{s}$, Eq.~(\ref{eq:singlecoeff}) becomes
\begin{eqnarray}
A_0 & = & 2\pi w_0sA_0 \nonumber \\
A_1 & = & \pi w_1sA_1 \label{A011}
\end{eqnarray}
where $|A_0| \ge A_1$, so that $U(x) > 0$ for all $x$.  We additionally require that $A_1>0,$ so that the peak of the bump corresponds to the stored spatial position. To ensure that a continuum of values of $(A_0, A_1)$ exist that solve Eq.~(\ref{A011}) in this case, we require that $s = \frac{1}{\pi w_1}$ and $w_1=2w_0$.  Thus, excitation $w_1$ must properly balance inhibition and the tuning of the firing rate gain $s$ must be inversely proportional to excitation for proper parameter balance.

\begin{figure}
\begin{center}
\includegraphics[scale=1]{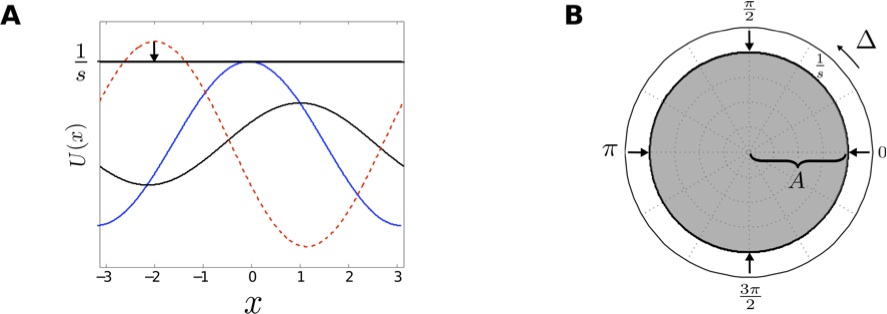}
\end{center}
\caption{Stationary Bump Solutions.  (\textbf{A}) Blue and black indicate neutrally stable solutions for $U(x)$ while red indicates unstable solutions that are attracted to the boundary of the line attractor.  (\textbf{B}) Polar plot of neutrally stable region, with the amplitude $A$ as the radial parameter and $\Delta$ as phase parameter.  The grey area shows where solutions of the form $A\cos{(x-\Delta)}$ are neutrally stable.}\label{fig:neutralsingle}
\end{figure}

Next, we will examine solutions that obey the restriction $U(x) \le \frac{1}{s}$, but have some values at which $U(x)<0$. This corresponds to bumps in which only a portion of the population is active. Again, we will show that the recurrent excitation in the network must be properly balanced by the gain of the system to yield a continuum of allowable bump amplitudes. From our analysis in section \ref{model}, we can conclude that if the synaptic weight function $w$ is unimodal, then $U(x)$ may be unimodal as well and the solution will exhibit two roots, $U( \pm a) = 0$. Then the system of equations (\ref{eq:singlecoeff}) becomes
\begin{eqnarray}
A_0 & = & 2sw_0\left[aA_0+A_1\sin{a}\right] \nonumber \\
A_1 & = & sw_1\left[A_0\sin{a}+aA_1\right] \label{A012}.
\end{eqnarray}
Requiring self-consistency implies that either $A_1 = A_0$ or $A_0 = 0$ and $A_1 < \frac{1}{s}$.  Therefore, we must either have that
\begin{equation}  \label{eq:singpos}
U(x) = A(1+\cos{x}),
\end{equation}
or
\begin{equation} \label{eq:singlesol}
U(x)  =  A\cos{x}.
\end{equation}
Since equation (\ref{eq:singpos}) satisfies the more restrictive condition $0 \leq U(x) \leq \frac{1}{s}$ that we have already considered, we will focus on bump solutions of the form given by Eq.~(\ref{eq:singlesol}). To obtain solutions of this form, we must have $w_0 =0$.  Now using Eq.~(\ref{eq:singlecoeff}) we find that
\begin{equation*}
A = \begin{cases}
sAw_1\left[\frac{\pi}{2}-\cos^{-1}\left(\frac{1}{sA}\right)-\frac{1}{sA}\sqrt{1-\left(\frac{1}{sA}\right)^2}\right] + 2w_1\sqrt{1-\left(\frac{1}{sA}\right)^2}, & \text{\indent for $sA > 1$}, \\
sAw_1\frac{\pi}{2}, & \text{\indent for $sA \le 1$},
\end{cases}
\end{equation*}
which requires that $w_1 = \frac{\D 2}{\D \pi s}$.  This means that recurrent excitation must be inversely proportional to the gain of the system.  When $sA \le 1$ we have a continuum of solutions for $A \in [0,\frac{1}{s}]$ that are all stationary solutions. When $sA >1$ there is no solution for $A$ for the given fixed value of $s$.  Thus the only fixed points are in the interval $[0, \frac{1}{s}]$ as illustrated in Fig. \ref{fig:neutralsingle}A.  As we will show in the next section, any amplitude above the threshold will decay back down to the boundary of the line attractor (red dashed plot in Fig. \ref{fig:neutralsingle}A).  We also illustrate in Fig. \ref{fig:neutralsingle}B, and show in the stability analysis, that any phase shifted bump will be neutrally stable as well.  Thus we have a two-dimensional attractor surface on the closed disc of radius $\frac{1}{s}$ each point of which corresponds to a bump solution.

\subsection{Stability of Bump Solution}\label{sec:stability1pop}
To show that the solutions we described above are neutrally stable, we carry out a linear stability analysis of the stationary bump solutions $U(x)$ of Eq.~(\ref{eq:singlenet}).  In particular, we study the temporal evolution of small, smooth, separable perturbations $\e^{\lambda t} \psi (x)$ by plugging in the linear expansion
\begin{equation}
 u(x,t) = U(x) + \psi(x)e^{\lambda t}  \label{ulinz}
 \end{equation}
 where $||\psi(x)||\ll1$. Since $\psi(x)$ must be periodic, we expand it in a Fourier series
 \begin{equation}  \label{psiexp}
 \psi(x) = \sum_{k=1}^{N}\A_k\cos{kx}+\sum_{k=1}^N\B_k\sin{kx}.
 \end{equation}
The perturbed bump solution given in Eq.~(\ref{ulinz}) can then be plugged into Eq.~(\ref{eq:singlenet}) and linearized to yield the general spectral problem \citep{ermentrout98,veltz10} 
 \begin{align}
 ( \lambda + 1) \psi (x) = \int_{- \pi}^{\pi} w(x,y) f'(U(y)) \psi (y) \d y  \label{geneprob}
 \end{align}
 for the stability of the bump. The associated coefficients of the expansion in Eq.~(\ref{psiexp}) are then determined by the linear system
\begin{align}
 {\mc A}_k = w_k \int_{- \pi}^{\pi} \cos (kx) f'(U(x)) \psi (x) \d x, \ \ \ {\mc B}_l = w_l \int_{- \pi}^{\pi} \sin (lx) f'(U(x)) \psi (x) \d x,
\end{align}
where $k,l = 1,...,N$. Solutions of this system, along with the associated $\lambda,$ are eigensolutions of Eq.~(\ref{geneprob}). We can directly compute the eigenvalues associated with the stability of bumps in the case of the weight function in Eq.~(\ref{eq:singleweight}) so that
 \begin{equation}
 (\lambda+1)\psi(x) = w_1\int_{-\pi}^{\pi}\cos{(x-y)}f'(U(y))\psi(y)dy. \label{linlam}
 \end{equation}
Analyzing solutions $(\lambda, \psi)$ of Eq.~(\ref{linlam}) is equivalent to determining the elements of the spectrum of the linear system in the vicinity of the bump. We are mainly interested in the point spectrum of the linear operator in Eq.~(\ref{linlam}), since the sign of the real part of $\lambda$ for these solutions will determine the associated stability of stationary bump solutions (see \citep{coombes04,veltz10} for detailed discussions of the partitioning of spectra in neural field models). In particular, we  examine the stability of stationary bump solutions of the form $U(x) = A \cos x$ when the firing rate function has the form given in Eq.~(\ref{eq:firingrate}).  Hence,
 \begin{eqnarray}
 (\lambda+1)\A_1 & = & 
 \begin{cases}
 \A_1sw_1\left[\frac{\pi}{2}-\cos^{-1}\left(\frac{1}{sA}\right)-\frac{1}{sA}\sqrt{1-\left(\frac{1}{sA}\right)^2}\right], & \text{\indent for $sA > 1$}, \\
 \A_1sw_1\frac{\pi}{2}, & \text{\indent for $sA \le 1$},
 \end{cases} \label{slam1} \\
 (\lambda+1)\B_1 & = & 
 \begin{cases}
 \B_1sw_1\left[\frac{\pi}{2}-\cos^{-1}\left(\frac{1}{sA}\right)+\frac{1}{sA}\sqrt{1-\left(\frac{1}{sA}\right)^2}\right], & \text{\indent for $sA > 1$} ,\\
 \B_1sw_1\frac{\pi}{2}, & \text{\indent for $sA \le 1$},
 \end{cases}  \label{slam2}
 \end{eqnarray}
and $\A_k = \B_k = 0$ for $k \ne 1$ and $\A_0 = 0$.  Now, bump solutions of Eq.~(\ref{eq:singlenet}) will be neutrally stable to both even and odd perturbations when parameters in Eqs.(\ref{slam1}--\ref{slam2}) are such that some solutions have ${\rm Re} \lambda = 0$ and others have  ${\rm Re}\lambda < 0$. 

We show that for the conditions we derived in the previous section, when $A \in [0, \frac{1}{s}]$, this is the case.  It is easy to see that if $A \le \frac{1}{s}$, we obtain $\lambda = 0$ corresponding to both even and odd perturbations by requiring that $w_1 = \frac{2}{\pi s}$ as was the condition for finding the stationary amplitudes in section 3.1.  When $A > \frac{1}{s}$, we find
\begin{eqnarray}
\lambda_o & = & \frac{2}{\pi}\left(-\cos^{-1}\left(\frac{1}{sA}\right)-\frac{1}{sA}\sqrt{1-\left(\frac{1}{sA}\right)^2}\right) < 0, \\
\lambda_e & = & \frac{2}{\pi}\left(-\cos^{-1}\left(\frac{1}{sA}\right)+\frac{1}{sA}\sqrt{1-\left(\frac{1}{sA}\right)^2}\right) < 0,
\end{eqnarray}
so that the bump is linearly stable in this region.  Thus if the amplitude of the bump exceeds this threshold, it will decay back down to $\frac{1}{s}$. Therefore, the saturating threshold $1/s$ sets an upper limit on the certainty that can be encoded by the amplitude $A$. The two-dimensional surface of hence consists of neutrally stable bumps to which all solutions are attracted.

\subsection{Integrating Input}

In an oculomotor delayed response task, an observer is presented with a spatial cue, for a time period $T_0$, during which the position of the cue must be remembered.  During this period, the cue location is encoded in the network activity by integration of the stimulus.  By the end of the presentation a bump of activity arises setting the initial condition for the solution during the delay period.  We next propose a way of determining how the cue is integrated by the network.  More specifically we study how the amplitude of the bump evolves during this integration period and how it evolves during the delay period, once the cue disappears.  We assume the stimulus current has the form

\begin{equation}\label{current}
I(t) = I_0(t)\left(H(t)-H(t-T_0)\right)\cos{x},
\end{equation}
so that input starts at $t=0$, ends at $t = T_0$ and has magnitude $I_0(t)$.  We write the solution to Eq.~(\ref{eq:singlenet}) in the form 
\begin{equation*}
u(x,t) = A(t)\cos{x}.
\end{equation*}
Substituting this into Eq.~(\ref{eq:singlenet}) we find
\begin{equation}\label{eq:Adot}
\dot{A}(t) = 
\begin{cases}
I_0(t), & \text{\indent if $T_0 \le t_{max}$}, \\
g(t), & \text{\indent if $T_0 > t_{max}$},
\end{cases}
\end{equation}
where $g(t)$ is the piecewise function
\begin{equation}
g(t) = 
\begin{cases}
I_0(t), & \text{\indent if $t < t_{max}$}, \\
I_0(t) - sA(t)w_1\left[\cos^{-1}\left(\frac{1}{sA(t)}\right) + \frac{1}{sA(t)}\sqrt{1-\left(\frac{1}{sA(t)}\right)^2}\right] + 2w_1\sqrt{1-\left(\frac{1}{sA(t)}\right)^2}, & \text{\indent if $t > t_{max}$},
\end{cases}
\end{equation}
and $t_{max}$ is the time at which $A(t) = \frac{1}{s}$.  We see that, in the line attractor region, the network simply integrates the amplitude of $I(t)$. Numerical methods must be employed to solve for $A(t)$ when it exceeds this value.  However for amplitudes beyond this range, solutions are attracted toward the boundary of the line attractor,  and hence
\begin{equation}
\lim_{t \to \infty}A(t) = 
\begin{cases}
\int_0^{T_0}I_0(t), & \text{\indent if $T_0 \le t_{max}$}, \\
\frac{\pi w_1}{2}, & \text{\indent if $T_0 > t_{max}$}.
\end{cases}
\end{equation}

Typically, in an experiment, an observer would be exposed to the spatial cue for a short time, $T_0$. Furthermore the cue would be at full contrast instantly.  We therefore set $I_0(t) \equiv I_0$ to be constant, and find from Eq.~(\ref{eq:Adot}) that if $T_0 \le t_{max}$ then 
\begin{equation}
A(t) = 
\begin{cases}
I_0t, & \text{for $t \le T_0$}, \\
I_0T_0, & \text{for $t > T_0$}.
\end{cases}
\end{equation}
It is easy to find that 
\begin{equation*}
t_{max} = \frac{1}{sI_0}
\end{equation*}
by solving $A(t_{max}) = \frac{1}{s}$. \\  

\begin{figure}
\begin{center}
\includegraphics[scale=1]{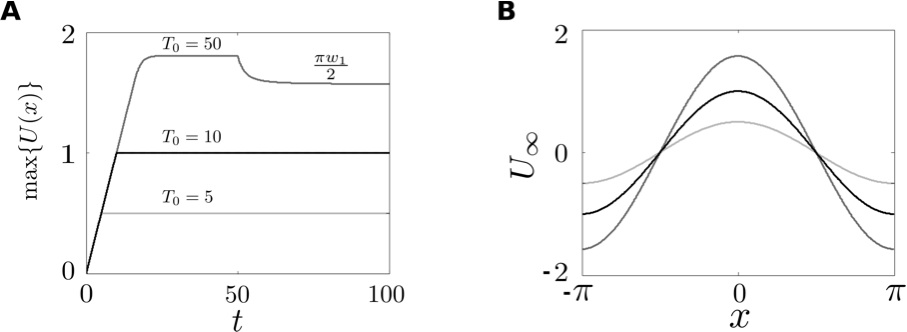}
\end{center}
\caption{The integration of constant external input $I(t) = I_0$ lasting for $T_0$ time units by the single population network (\ref{eq:singlenet}).  (\textbf{A}) The amplitude of the bump in response to a current injection of the form (\ref{current}). (\textbf{B}) Bump solution as $t \to \infty$.}\label{fig:intamp}
\end{figure}

In general, the integration of the spatial cue has a very important consequence in terms of encoding certainty.  As mentioned above, experimental evidence suggests that higher uncertainty is preceded by lower neuronal activity and vice versa.  Thus we show here that in our model, longer cue exposure leads to greater bump amplitudes which we interpret as greater initial certainty.  Additionally, greater cue contrast, $I_0$, will also lead to larger bump amplitudes.  Therefore we can reach the same bump amplitudes for shorter cue times by increasing the contrast of the cue.  It is also important to note that the amplitude eventually saturates to a maximum value, even in the presence of infinite cue time (see Fig.~\ref{fig:intamp}A).  Therefore we always have an upper bound on the amount of transient initial certainty, corresponding to transient high values of bump amplitude.  Once the cue is turned off the amplitude relaxes to the boundary of the line attractor set by the saturating threshold $1/s$ and corresponding to the maximal long-term certainty of the system.

\subsection{Diffusion of Bump in a Single Population Network}

Cortical neurons {\em in vivo} typically have high variability in their spike train output \citep{softky93}, arising from channel noise \citep{white00} as well as a high level of background synaptic input not linked to a circuit's immediate task \citep{faisal08}. Effects of these fluctuations are typically incorporated into neural field models by considering finite sized corrections to the mean field \citep{ginzburg94,elboustani09,bressloff09}. Truncating stochastic terms to linear order then yields Langevin equations that can be analyzed using asymptotic techniques for stochastic partial differential equations \citep{hutt08,bressloff12}. Here, we consider a phenomenological model that incorporates fluctuations into a neural field model, which has recently been used to study the effects of noise on spatiotemporal patterns \citep{hutt08,bressloff12b,kilpatrick13}
\begin{equation} \label{eq:singlestoch}
d\U (x,t) = \left[-\U (x,t) + \int_{-\pi}^{\pi}w(x-y)f(\U (y,t)) \d y  \right]dt + \varepsilon^{1/2}dW(x,t),
\end{equation}
where
\begin{equation}\label{eq:dW}
\langle dW(w,t) \rangle = 0, \text{\indent} \langle dW(x,t)dW(y,s) \rangle = C(x-y)\delta(t-s)dtds,
\end{equation}
and $\varepsilon \ll1$ is the amplitude of noise. In our analysis, we use an ansatz originally used to study the effects of noise on wave propagation in stochastic reaction-diffusion equations \citep{armero98}. In line with previous studies \citep{armero98,bressloff12b,kilpatrick13}, we assume that the noise term leads to diffusion of the bump's position. In addition, due to the additional neutrally stable direction associated with the bump's amplitude, we assume the amplitude will diffuse in response to noise too.  With this in mind we can express the solution,  $\U,$ as the sum of a fixed bump profile, $U$, shifted in its phase by $\Delta(t)$ (which represents the remembered position), increased/decreased in amplitude by amount $\xi(t)$, and higher order time-dependent fluctuations $\varepsilon^{1/2}\Phi + \varepsilon\Phi_1+\varepsilon^{3/2}\Phi_2+...$ in the profile of the bump. Hence, we write
\begin{equation}
\U (x,t) = (1+\xi_A(t))U(x-\Delta(t)) + \varepsilon^{1/2}\Phi(x-\Delta(t),t),
\end{equation}
where, for convenience, we use the normalized stochastic variable
\begin{equation}
\xi_A(t) = \frac{\xi(t)}{A_0}.
\end{equation}
Plugging this into Eq.~(\ref{eq:singlestoch}) we get
\begin{eqnarray}\label{eq:dphi}
d\Phi(x,t) & = & \varepsilon^{-1/2}U'(x)d\Delta(t) + \mathcal{L}\Phi(x,t) - \varepsilon^{-1/2}d\xi_A(t)U(x)\nonumber \\
& & +\varepsilon^{-1/2}\xi_A(t)\mathcal{L}U(x)+\varepsilon^{-1/2}\xi_A(t)U'(x)d\Delta(t) + dW(x,t),
\end{eqnarray}
where
\begin{equation}
\mathcal{L}p(x) = -p(x) + \int_{-\pi}^{\pi}w(x-y)f'(U(y))p(y)dy. \label{linop}
\end{equation}
In the case of the weight function given by Eq.~(\ref{eq:singleweight}) tuned so that the amplitude of the bump is neutrally stable, we also have that $\mathcal{L}U(x) = 0$.  Then Eq.~(\ref{eq:dphi}) can be rewritten as
\begin{equation}\label{eq:dphired}
d\Phi(x,t) = \varepsilon^{-1/2}U'(x)(1+\xi_A(t))d\Delta(t) + \mathcal{L}\Phi(x,t) - \varepsilon^{-1/2}d\xi_A(t)U(x) + dW(x,t).
\end{equation}
We can ensure that a bounded solution exists by requiring that the inhomogeneous part of Eq. (\ref{eq:dphired}) be orthogonal to all elements of the nullspace of the adjoint operator $\mathcal{L}^{*}$, \citep{bressloff01,kilpatrick13}, where 
\begin{equation}\label{eq:adjoint}
\mathcal{L}^{*}q(x) = -q(x) + f'(U(x))\int_{-\pi}^{\pi}w(x-y)q(y)dy
\end{equation} 
Therefore, the equation defining the elements $\varphi(x)$ of the nullspace of ${\mc L}^*$ is
\begin{align}
\varphi(x) = f'(U(x)) \int_{-\pi}^{\pi} w(x-y) q(y) \d y.  \label{anul}
\end{align}
To identify the nullspace elements of ${\mc L}^*$, recall that we have required neutral stability ($\lambda =0$) with respect to the linear operator ${\mc L}$, defined in Eq.~(\ref{linop}), for an odd $\phi_o(x)$ and even $\phi_e(x)$ eigenfunction, so
\begin{align}
\phi_j (x) = \int_{-\pi}^{\pi} w(x-y) f'(U(y)) \phi_j(y) \d y, \ \ \ j=o,e. \label{lnul}
\end{align}
Setting $\varphi_j(x) = f'(U(x)) \phi_j(x)$ ($j=o,e$) in Eq.~(\ref{anul}), we have
\begin{align}
\varphi_j(x) = f'(U(x)) \phi_j(x) = f'(U(x)) \int_{- \pi}^{\pi} w(x-y) f'(U(y)) \phi_j(y)) \d y, \ \ \ j=o,e,
\end{align}
which holds according to Eq.~(\ref{lnul}). Thus, there are two functions that span the nullspace of $\mathcal{L}^{*}$: one even function, $\varphi_e(x)$, and one odd function, $\varphi_o(x)$. Taking the inner product of both sides of Eq. (\ref{eq:dphired}) with respect to each of these functions yields the following equations:
\begin{eqnarray}
\int_{-\pi}^{\pi}\varphi_o(x)\left[U'(x)(1-\xi_A(t))d\Delta(t) + \varepsilon^{1/2}dW(x,t)\right]dx & = & 0, \nonumber \\
\int_{-\pi}^{\pi}\varphi_e(x)\left[-U(x)d\xi_A(t) + \varepsilon^{1/2}dW(x,t)\right]dx & = & 0,
\end{eqnarray}
since $U(x)$ is even and $U'(x)$ is odd.  Solving for $d\Delta(t)$ and $d\xi_A(t)$, we find that
\begin{eqnarray}\label{eq:langeq}
d\Delta(t) & = & -\frac{\varepsilon^{1/2}}{1 + \xi_A(t)}\frac{\int_{-\pi}^{\pi}\varphi_o(x)dW(x,t)dx}{\int_{-\pi}^{\pi}\varphi_o(x)U'(x)dx}, \\
d\xi_A(t) & = & \varepsilon^{1/2}\frac{\int_{-\pi}^{\pi}\varphi_e(x)dW(x,t)dx}{\int_{-\pi}^{\pi}\varphi_e(x)U(x)dx}, \label{eq:dxi}
\end{eqnarray}
and we see that the stochastic variable $\Delta(t)$ depends on $\xi(t)$.  Therefore $\Delta(t)$ will not undergo linear diffusion. 

We proceed by first computing the distribution of $\xi(t)$, then and use it to find the distribution of $\Delta(t)$. Since $\langle \xi(t) \rangle = 0$ (the additive noise we apply is white in time), computing the variance of $\xi(t)$ we find that it evolves according to pure diffusion since
\begin{eqnarray}\label{eq:varxi1}
\langle \xi(t)^2 \rangle & = & \varepsilon A_0^2\frac{\int_{-\pi}^{\pi}\int_{-\pi}^{\pi}\varphi_e(x)\varphi_e(y)\langle W(x,t)W(y,t)\rangle dxdy}{\left[\int_{-\pi}^{\pi}\varphi_e(x)U(x)dx\right]^2} \, t,  \nonumber \\
& = & D_{\xi}(\varepsilon)t
\end{eqnarray} 
Using Eq.~(\ref{eq:dW}) we write the diffusion coefficient as
\begin{equation}
D_{\xi}(\varepsilon) = \varepsilon A_0^2\frac{\int_{-\pi}^{\pi}\int_{-\pi}^{\pi}\varphi_e(x)\varphi_e(y)C(x-y)dxdy}{\left[\int_{-\pi}^{\pi}\varphi_e(x)U(x)dx\right]^2}.
\end{equation}
Note that since $\mathcal{L}^*\varphi_e(x) = \mathcal{L}^*\varphi_o(x)= 0$, we use Eq.~(\ref{eq:adjoint}) and the general weight function given by Eq.~(\ref{eq:expandweight}) to find that
\begin{eqnarray}
\varphi_o(x) & = & f'(U(x)\sum_{k=1}^N\mathcal{S}_k\sin{(kx)}, \nonumber \\
\varphi_e(x) & = & f'(U(x)\sum_{k=1}^N\mathcal{C}_k\cos{(kx)},
\end{eqnarray}
where
\begin{eqnarray}
\mathcal{S}_k & = & w_k\int_{-\pi}^{\pi}\sin{(kx)}\varphi_o(x), \nonumber \\
\mathcal{C}_k & = & w_k\int_{-\pi}^{\pi}\cos{(kx)}\varphi_e(x).
\end{eqnarray}
This system can in general be solved using methods of linear algebra, \citep{veltz10,kilpatrick13}.  For the weight function $w(x,y) = w_1\cos{(x-y)}$ the system simplifies to
\begin{eqnarray}
\varphi_o(x) & = & w_1\mathcal{S}f'(A_0\cos{x})\sin{x}, \nonumber \\
\varphi_e(x) & = & w_1\mathcal{C}f'(A_0\cos{x})\cos{x}.
\end{eqnarray}
Therefore, we can solve Eq.~(\ref{eq:varxi1}) by first computing the integral in the denominator as
\begin{equation}
\int_{-\pi}^{\pi}\varphi_e(x)U(x)dx = w_1\mathcal{C}A_0\int_{-\pi}^{\pi}f'(A_0\cos{x})\cos^2{x}dx = \frac{\pi}{2}sw_1\mathcal{C}A_0,
\end{equation}
where we impose the condition for neutral stability, $s = \frac{2}{\pi w_1}$.  Thus, Eq.~(\ref{eq:varxi1}) becomes 
\begin{equation}
D_{\xi}(\varepsilon) = \varepsilon\int_{-\pi}^{\pi}\int_{-\pi}^{\pi} \cos{x}\cos{y}f'(U(x))f'(U(y))C(x-y) dxdy.
\end{equation}
We can approximate $\Delta(t)$ by expanding the $\xi_A(t)$ dependent term in Eq. (\ref{eq:langeq}) to second order, so
\begin{equation}
d\Delta(t) = -\varepsilon^{1/2}\left(1-\xi_A(t)+\xi_A(t)^2\right)\frac{\int_{-\pi}^{\pi}\varphi_o(x)dW(x,t)dx}{\int_{-\pi}^{\pi}\varphi_o(x)U'(x)dx},
\end{equation}
where we know that $\langle \xi_A(t) \rangle = 0$ and $\langle \xi_A(t)^2 \rangle = \frac{1}{A_0^2}\langle \xi(t)^2 \rangle $.  For short timescales we can approximate $\Delta(t)$ as a pure diffusion process by ignoring the $\xi_A(t)$ terms.  We proceed similarly to find $D_{\xi}(\varepsilon)$ for unimodal solutions  $U(x)$. Noting that $U'(x) = -A_0\sin{x}$, we find that 
\begin{equation}
D_{\Delta}(\varepsilon) = \frac{\varepsilon}{A_0^2}\int_{-\pi}^{\pi}\int_{-\pi}^{\pi}\sin{x}\sin{y}f'(U(x))f'(U(y))C(x-y)dxdy.
\end{equation}
Thus the diffusion coefficient decreases as the inverse square of the initial bump amplitude.  Since we interpreted this amplitude as a measure of certainty, this implies that the greater the certainty of the stored position the less the position diffuses during the delay period.  We have shown that the initial amplitude can be controlled by the duration and contrast at which the spatial cue presented to an observer.  Therefore longer exposure times along with contrast will determine the accuracy of recall. 

\subsection{Calculating effective stochastic motion of bumps}

\begin{figure}[hbp!]
\begin{center}
\includegraphics[scale=1]{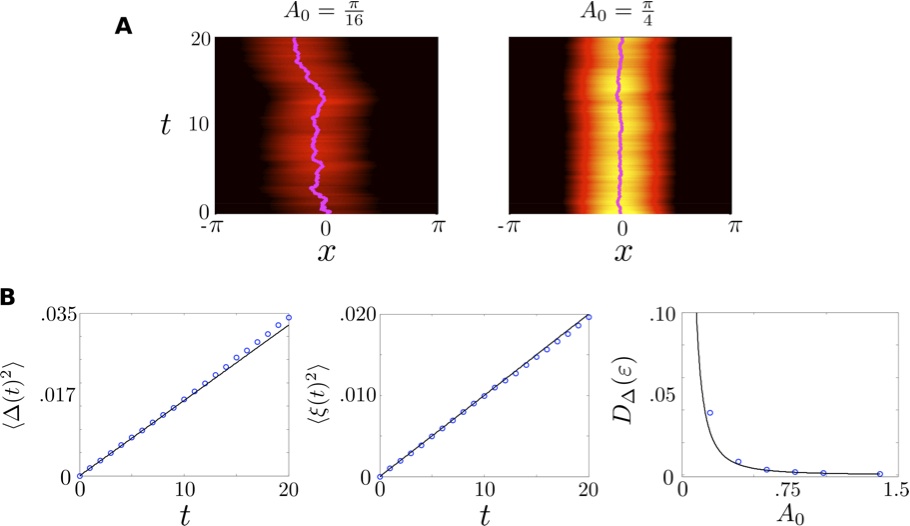}
\caption{Diffusion of Bump Solution. (\textbf{A}) Space-time plot of bump during the delay period for both low and high initial amplitude.  The bump's position diffuses more for smaller initial bump amplitude.  (\textbf{B})  Variance of the bump center, $\Delta(t)$, and bump amplitude $\xi(t)$ as a function of time as well as the the diffusion coefficient for $\Delta(t)$ as a function of initial amplitude, $A_0$.  Other parameters used: $\varepsilon = .001$, $s = 2/\pi$, $w_1 = 1$.  For the variance plots, $A_0 = \pi/4$}\label{fig:diffusion_lowe}
\end{center}
\end{figure}

We now compute the diffusion coefficients using cosine shaped spatial correlations, $C(x-y) = \cos{(x-y)}$.  Under this assumption, we find that for $A_0 \in \left[0, \frac{1}{s}\right]$
\begin{eqnarray}
D_{\xi}(\varepsilon) = \varepsilon \nonumber, \ \ \ \ \ D_{\Delta}(\varepsilon) =  \frac{\varepsilon}{A_0^2}.  \label{sdiffcos}
\end{eqnarray} 
As demonstrated by comparing single realizations of the stochastic equation (\ref{eq:langeq}) in Fig.~\ref{fig:diffusion_lowe}A, bumps with initially smaller amplitudes ($A_0= \pi/16$) diffuse more than bumps with larger amplitudes ($A=\pi/4$). In fact, we can see that both the relationship predicted by Eq.~(\ref{sdiffcos}) and simulations show that the diffusion of the bump's position decreases as the initial amplitude increases (Fig. \ref{fig:diffusion_lowe}B). Theory and simulations agree well, for noise amplitude of $\varepsilon = 0.001$. To account for dynamics occurring for larger values of $\varepsilon$ or longer timescales, we consider nonlinearities in Eq.~(\ref{eq:langeq}).  Additionally we must pay special attention to deriving the effective stochastic differential equation for $\xi(t)$, as a linear truncation becomes insufficient. 

\begin{figure}[hbp!]
\begin{center}
\includegraphics[scale=1]{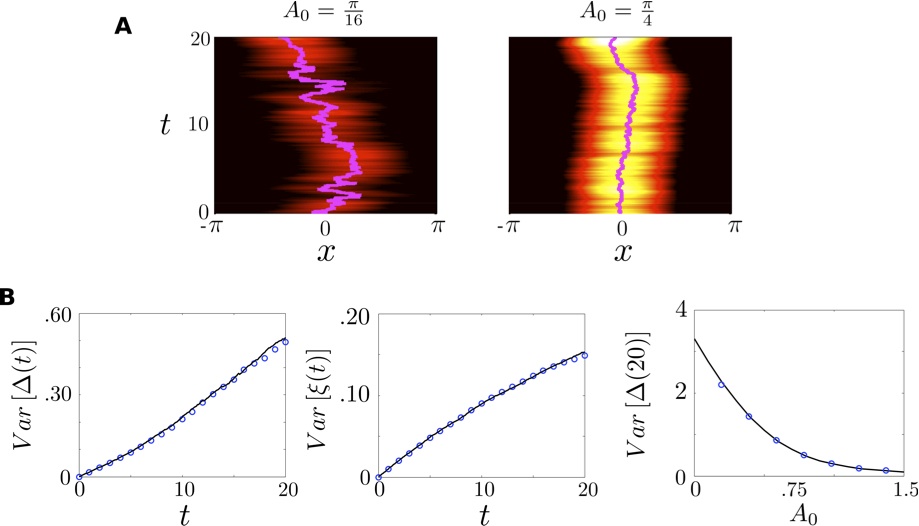}
\caption{Diffusion of Bump Solution. (\textbf{A}) Space-time plot of bump during the delay period for both low and high initial amplitude.  The bump's position diffuses more for smaller initial bump amplitude.  (\textbf{B})  Variance of the bump center, $\Delta(t)$, and bump amplitude $\xi(t)$ as a function of time as well as the the variance for $\Delta(t)$, evaluated at $t=20$, as a function of initial amplitude, $A_0$.  Other parameters used: $\varepsilon = .01$, $s = 2/\pi$, $w_1 = 1$.  For the variance plots, $A_0 = \pi/4$.}\label{fig:diffusion_highe}
\end{center}
\end{figure}

To derive a more accurate approximation of the variances of $\Delta(t)$ and $\xi(t)$ in the case of cosine shaped spatial noise correlations, we propose a more precise ansatz for the stochastic motion of the bump
\begin{equation}\label{eq:cosine_ansatz}
\U(x,t) = U_0(x) + A_1(t)\cos{x}+A_2(t)\sin{x}
\end{equation} 
where both $A_1(t), A_2(t)$ are stochastic variables.  Using the trigonometric identity for the sum of a sine and a cosine and the initial condition, $U_0(x) = A_0\cos(x)$, we can write 
\begin{equation}
\U(x,t) = \sqrt{(A_0+A_1(t))^2+A_2(t)^2}\cos{(x-\Delta(t))}
\end{equation}
where $\Delta(t) = \tan^{-1}\left(\frac{A_2(t)}{A_0+A_1(t)}\right)$. We use the equality, $\xi(t) = \sqrt{(A_0+A_1(t))^2+A_2(t)^2}-A_0$, to track the stochastic variable that measures the displacement of the amplitude from it's initial point. Therefore as long as this effective amplitude is in the interval $\left[0, \frac{1}{s}\right]$, then 
\begin{equation}
\U(x,t) = \int_{-\pi}^{\pi}w(x-y)f(\U(y,t))dy,
\end{equation}
is satisfied by the condition of neutral stability.  Therefore Eq.~(\ref{eq:singlestoch}) simplifies to 
\begin{equation}
dA_1(t)\cos{x}+dA_2(t)\sin{x} = \varepsilon^{1/2}(dW_1(t)\cos{x} + dW_2(t)\sin{x}),
\end{equation}
which we rewrite as
\begin{eqnarray}
dA_1(t) & = & \varepsilon^{1/2}dW_1(t), \nonumber \\
dA_2(t) & = & \varepsilon^{1/2}dW_2(t).
\end{eqnarray}
We see that this is equivalent to a 2D diffusion process with initial condition, $(A_1(0), A_2(0)) = (0, 0)$ and
\begin{eqnarray}
\langle A_1(t)^2 \rangle & = & \langle A_2(t)^2 \rangle = \varepsilon t, \\ \nonumber
\langle A_1(t) \rangle & = & \langle A_2(t) \rangle = 0.
\end{eqnarray}
Therefore  to compute the variance of the amplitude we must compute
\begin{eqnarray}
\langle \xi(t)^2 \rangle - \langle \xi(t) \rangle^2 & = & A_0^2 + 2\varepsilon t - \langle \sqrt{(A_0+A_1(t))^2+A_2(t)^2} \rangle^2,
\end{eqnarray}
where the last term must be determined using Monte Carlo simulations.
Again, we obtain the relationship that $\Delta(t)$ decreases with increasing initial bump amplitude, $A_0$. As shown in Fig.~(\ref{fig:diffusion_highe}B) the Monte Carlo simulation of the stochastic process $\Delta(t)$ agrees well with the full system.  Additionally, we see that the variance of $\Delta(t)$ indeed does decrease with initial amplitude, $A_0$. Thus, the certainty of bumps relates to how sensitive they will be to stochastic fluctuations. More initial certainty (higher $A_0$) translates to a lower diffusion coefficient across a broad range of noise amplitudes.

\section{Obtaining neutral stability through excitatory-inhibitory balance}
\label{einet}

So far we have considered a network described by Eq.~(\ref{eq:singleweight}) which lumps excitatory and inhibitory cells into a single population. In doing so, we were able to restrict the set of parameters to generate a network which supported a set of bump solutions with a continuum of amplitudes. We can perform a similar analysis in a network with two separate populations for excitatory and inhibitory cells, but our restrictions on parameters simply involve more conditions. As explained in section 2, the single population network is equivalent to the two population network in the limit of fast inhibition, $\tau \to 0$.  However, as  inhibition becomes slower, $\tau > 0$, it is no longer clear that stability in the single population implies stability in the two population network.  We will see that the conditions for finding a continuum of fixed points are exactly the same as in the single population, since the stationary solutions do not depend on $\tau$. However the equations used in the stability analysis do depend on $\tau$. Excessively slow inhibition can destabilize stationary bump solutions. Perturbations that translate the position of the bump are always neutrally stable, due to the underlying translation symmetry of the network \citep{bressloff01,kilpatrick13}. However, generating a network with bumps that are neutrally stable to even symmetric perturbations depends on conditions that relate to the speed of inhibition.

\subsection{Stationary Bump Solution}
We now study the excitatory-inhibitory network defined by Eq.~(\ref{eq:einet}) with synaptic weights determined by the functions in Eq.~(\ref{eq:multiweight}).   First we look for even stationary bump solutions of the form
\begin{eqnarray}\label{eq:gensol_ei}
u(x,t) =& U(x) = A_0 + A_1\cos{x}, \nonumber \\
v(x,t) =& V(x) = M_0 + M_1\cos{x}.
\end{eqnarray}
Note this ansatz implies time-derivatives $u_t = v_t = 0$ in Eq.~(\ref{eq:einet}). Thus, by substituting in $V(x)$ equation into the $u$ equation we can generate the single equation
\begin{equation}\label{eq:Ufixed_ei}
U(x) = (w_{ee}(x)-w_{ei}(x)*w_{ie}(x))*f(U(x)),
\end{equation}
where $f(x)*g(x) = \int_{-\pi}^{\pi}f(x-y)g(y)dy$. Therefore, stationary solutions to Eq.~(\ref{eq:einet}) are the same as stationary solutions to Eq.~(\ref{eq:singlenet}), under the requirement that the assignment of the effective synaptic weight function
\begin{equation}\label{eq:eff_weight}
w(x) = w_{ee}(x)-w_{ei}(x)*w_{ie}(x) = \wee-2\pi\wei\wie+\wee\cos{x}.
\end{equation}
Note that Eq.~(\ref{eq:eff_weight}) is equivalent to Eq.~(\ref{eq:singleweight}) by setting $w_0 = \wee-2\pi\wei\wie$ and $w_1 = \wee$.  Therefore, under an appropriate change of variables, solving Eq.~(\ref{eq:Ufixed_ei}) is equivalent to solving Eq.~(\ref{eq:Ufixed_single}). Therefore, our results concerning the existence of a continuum of amplitudes concerning Eq.~(\ref{eq:Ufixed_single}) should hold here as well. This means that in order to obtain a line attractor of bump amplitudes, we must have that $A_0=0$ and $\wee=2\pi\wei\wie$ (i.e $w_0 = 0$). However, we can still have $M_0 \ne 0$.  Additionally, analogous to the single network in Eq.~(\ref{eq:Ufixed_single}), we must require that $\wee = \frac{2}{\pi s}$.  Again, we have 
\begin{equation}\label{eq:Acontinuum}
A = \begin{cases}
sA\wee\left[\frac{\pi}{2}-\cos^{-1}\left(\frac{1}{sA}\right)-\frac{1}{sA}\sqrt{1-\left(\frac{1}{sA}\right)^2}\right]+2\wee\sqrt{1-\left(\frac{1}{sA}\right)^2}, & \text{\indent for $sA > 1$}, \\
sA\wee\frac{\pi}{2}, & \text{\indent for $sA \le 1$},
\end{cases}
\end{equation}
and, for the $v$ equation
\begin{eqnarray*}
V(x) = \wei\int_{-\pi}^{\pi}(1+\cos{(x-y)})f(U(y))dy,
\end{eqnarray*}
so that
\begin{eqnarray}
M_0 & = & 
\begin{cases}
2s\wei A\left[1-\sqrt{1-\left(\frac{1}{sA}\right)^2}\right]+2\wei\cos^{-1}\left(\frac{1}{sA}\right), & \text{\indent for $sA > 1$}, \\
2s\wei A, & \text{\indent for $sA \le 1$},
\end{cases} \nonumber \\
M_1 & = & \begin{cases}
sA\wei\left[\frac{\pi}{2}-\cos^{-1}\left(\frac{1}{sA}\right)-\frac{1}{sA}\sqrt{1-\left(\frac{1}{sA}\right)^2}\right]+2\wei\sqrt{1-\left(\frac{1}{sA}\right)^2}, & \text{\indent for $sA > 1$}, \\
sA\wei\frac{\pi}{2}, & \text{\indent for $sA \le 1$}.
\end{cases}
\end{eqnarray}
Again, we have a continuum of values for $A \in [0, \frac{\pi\wee}{2}]$ that are fixed points, and the coefficients for $v$ will depend on $A$, and upon substituting values for $s$ we obtain
\begin{equation}
M_0 = \frac{4\wie}{\pi\wee} A, \text{\indent} M_1 = \frac{\wie}{\wee}A.
\end{equation}

To study the way in which the line attractor globally organizes dynamics, we consider effects of breaking this balance condition in two ways: excess excitation or excess inhibition.  As we shall see, too much inhibition leads to no stable bump solutions whereas too much excitation leads to only a single stable bump solution. To do this, we define the quantity
\begin{equation}
\bar{w} = \wee - 2\pi\wei\wie
\end{equation}
and simply consider when $\bar{w}>0$ (excess excitation) and $\bar{w}<0$ (excess inhibition).

First let $\bar{w} < 0$ (excess inhibition) and consider when $U(x) < \frac{1}{s}$.  Then, similar to section \ref{stationary_single}, we find that 
\begin{eqnarray}\label{eq:A_equation1}
A_0 & = & 2s\bar{w}\left[aA_0+\sin{a}A_1\right] \nonumber \\
A_1 & = & s\wee\left[\sin{a}A_0+aA_1\right]
\end{eqnarray}
where $a = \cos^{-1}\left(-\frac{A_0}{A_1}\right)$ and $|A_0| \le |A_1|$.  We must consider the cases when $A_0>0$, $A_0 < 0$ and $A_0=0$.  If $A_0>0$, then since $0 \le a \le \pi$ we know that $\sin{a} \ge 0$.  Also, we impose that $A_1 \ge 0$ so that the peak of the bump corresponds to the remembered location of the stimulus.  Then, since $\bar{w}<0$, Eq.  (\ref{eq:A_equation1}) implies that $A_0$ equals something negative, which is a contradiction.  Now assume that $A_0 < 0$. Then Eq.~(\ref{eq:A_equation1}) implies that
\begin{equation*}
a  \ge  \frac{A_1}{|A_0|}\sin{a}, \qquad \text{and} \qquad
a  \ge  \frac{|A_0|}{A_1}\sin{a},
\end{equation*}
which implies that $|A_0| = A_1$.  Then our only choices are $U(x) = A_1(\cos{x}-1)$ or $U(x) \equiv 0$.  However, if the former were true, then $f(u) \equiv 0$ which forces $U(x) \equiv 0$.  Finally it is easy to see that if $A_0 =0$ then $A_1 = 0$ for $\bar{w} \ne 0$.

\begin{figure}[htp!]
\begin{center}
\includegraphics[scale=1]{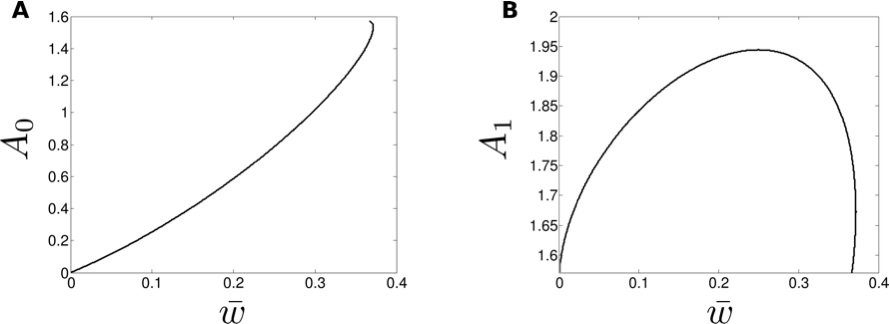}
\end{center}
\caption{Amplitudes of the stationary bump solutions $U(x) = A_0 + A_1\cos{x}$ for varying $\bar{w}$ values as per equation (\ref{eq:A_equation2}).  (\textbf{A}) $A_0$ as a function of $\bar{w}$. (\textbf{B}) $A_1$ as a function of $\bar{w}$.  Other values used: $s = \frac{2}{\pi}$, $\wee = 1$}\label{fig:Avsw}
\end{figure}

Now assume that $\bar{w} > 0$ (excess excitation).  In the case $U(x) < \frac{1}{s}$, we find that the only solution is $U(x) \equiv 0$. When $U(x) > \frac{1}{s}$ for some $x$, then equation (\ref{eq:A_equation1}) becomes
\begin{eqnarray}\label{eq:A_equation2}
A_0 & = & 2s\bar{w}\left[(a-b)A_0 + (\sin{a}-\sin{b})A_1\right] + 2\bar{w}b, \nonumber \\
A_1 & = & s\wee\left[(\sin{a}-\sin{b})A_0 + (a-b)A_1\right] + 2\wee\sin{b},
\end{eqnarray}
where $b = \cos^{-1}\left(\frac{1-sA_0}{sA_1}\right)$ such that $U(b) = \frac{1}{s}$.  To simplify the analysis, we will let $s = \frac{2}{\pi\wee}$ as was the condition for the line attractor.  We cannot solve the system in Eq.~(\ref{eq:A_equation2}) analytically, so we compute the solutions numerically and plot them in Fig.~\ref{fig:Avsw}.  We see that for any given $\bar{w}$ there is either only a single solution or two solutions for $U(x)$, therefore there are no line attractors in this case.  This means that the only case that yields a line attractor of bump amplitudes is when $\bar{w} = \bar{w}_{ee} - 2 \pi \bar{w}_{ei} \bar{w}_{ie} = 0$, when recurrent excitation is perfectly balanced by feedback inhibition. Thus, by considering separate excitatory and inhibitory populations, we see that we must place additional restrictions on the parameters of our model to attain a continuum of bump amplitudes. In the next section we will compute the stability of these solutions, showing we must consider the timescale $\tau$ of inhibition that feeds back upon the excitatory population.

\subsection{Stability of Bump Solution}\label{stabilityei}
We now perform a stability analysis on the fixed bump solution in Eq.~(\ref{eq:gensol_ei}).  We consider the set of parameters $A_0 =0$ and $A_1 = A \in [0, \frac{1}{s}]$ that leads to a line attractor of amplitudes. Similar to section \ref{sec:stability1pop}, we study the temporal evolution of small, smooth, separable perturbations, $e^{\lambda t}\psi(x)$ and $e^{\lambda t}\phi(x)$, by plugging in the linear expansion
\begin{eqnarray}\label{eq:uvexpand}
u(x,t) & = & U(x) + \psi(x)e^{\lambda t}, \nonumber \\
v(x,t) & = & V(x) + \phi(x)e^{\lambda t}, 
\end{eqnarray}
where $||\psi(x)||,||\phi(x)|| \ll 1$ and since both solutions must be periodic,
\begin{eqnarray}  \label{psphiser}
\psi(x) & = & \sum_{k=0}^N\A_k\cos{(kx)}+\sum_{k=1}^N\B_k\sin{(kx)}, \nonumber \\
\phi(x) & = & \sum_{k=0}^N\M_k\cos{(kx)}+\sum_{k=1}^N\N_k\sin{(kx)}.
\end{eqnarray}
Plugging the ansatz given by Eq.~(\ref{eq:uvexpand}) into Eq.~(\ref{eq:einet}) we obtain
\begin{eqnarray}
(\lambda+1)\psi(x) & = & w_{ee}*(f'(U(x))\psi(x))-w_{ie}*\phi(x), \nonumber \\
(\tau\lambda+1)\phi(x) & = & w_{ei}*(f'(U(x))\psi(x)). \label{psiphisys}
\end{eqnarray}
Similar to section \ref{single}, we analyze the solutions $(\lambda, \psi,\phi)$ to determine the stability of the perturbations by observing the sign of the real part of $\lambda$.  By self-consistency of (\ref{psphiser}) with (\ref{psiphisys}), we see that when using the weight functions in Eq.~(\ref{eq:multiweight}) we have the system
\begin{eqnarray}\label{eq:ei_coeff}
(\lambda + 1)\A_0 & = & \wee\int_{-\pi}^{\pi}\left(\A_0 + \A_1\cos{y}+\B_1\sin{y}\right)f'(U(y))dy-\wie\int_{-\pi}^{\pi}\left(\M_0+\M_1\cos{y}+\N_1\sin{y}\right)dy, \nonumber \\
(\lambda+1)\A_1 & = & \wee\int_{-\pi}^{\pi}\cos{y}\left(\A_0 + \A_1\cos{y}+\B_1\sin{y}\right)f'(U(y))dy, \nonumber \\
(\lambda+1)\B_1 & = & \wee\int_{-\pi}^{\pi}\sin{y}\left(\A_0 + \A_1\cos{y}+\B_1\sin{y}\right)f'(U(y))dy, \nonumber \\
(\tau\lambda+1)\M_0 & = & \wei\int_{-\pi}^{\pi}\left(\A_0 + \A_1\cos{y}+\B_1\sin{y}\right)f'(U(y))dy, \nonumber \\
(\tau\lambda+1)\M_1 & = & \wei\int_{-\pi}^{\pi}\cos{y}\left(\A_0 + \A_1\cos{y}+\B_1\sin{y}\right)f'(U(y))dy, \nonumber \\
(\tau\lambda+1)\N_1 & = & \wei\int_{-\pi}^{\pi}\sin{y}\left(\A_0 + \A_1\cos{y}+\B_1\sin{y}\right)f'(U(y))dy,
\end{eqnarray}
where $\A_k = \B_k$ for $k \ne 0,1$.  When $\tau \ne 0$, we can compute the integrals and set conditions of the parameters for the line attractor to find that the system in Eq.~(\ref{eq:ei_coeff}) is equivalent to the linear system
\begin{equation*}
\lambda \begin{pmatrix} \A_0 \\ \A_1 \\ \B_1 \\ \M_0 \\ \M_1 \\ \N_1 \end{pmatrix} = 
\begin{pmatrix}
1 & \frac{4}{\pi} & 0 & -2\pi\wie & 0 & 0 \\
\frac{4}{\pi} & 0 & 0 & 0 & 0 & 0 \\
0 & 0 & 0 & 0 & 0 & 0 \\
\frac{1}{\pi\wie\tau} & \frac{2}{\pi^2\wie\tau} & 0 & -\frac{1}{\tau} & 0 & 0 \\
\frac{2}{\pi^2\wie\tau} & \frac{1}{2\pi\wie\tau} & 0 & 0 & -\frac{1}{\tau} & 0 \\
0 & 0 & \frac{1}{2\pi\wie\tau} & 0 & 0 & -\frac{1}{\tau}
\end{pmatrix}
 \begin{pmatrix} \A_0 \\ \A_1 \\ \B_1 \\ \M_0 \\ \M_1 \\ \N_1 \end{pmatrix}.
 \end{equation*}
The associated matrix has the characteristic equation
\begin{equation}
\lambda^2(\tau\lambda+1)^2\left(\tau\lambda^2+(1-\tau)\lambda+1-\frac{16}{\pi^2}\tau\right)=0  \label{twopopchar}
\end{equation}
from which we obtain only two zero eigenvalues corresponding to odd perturbations $(0,0,1,0,0,\frac{\wei}{\wee})$ and even perturbations $(0,1,0,\frac{4\wei}{\pi\wee},\frac{\wei}{\wee},0)$.  Thus we see that obtaining a zero eigenvalue associated with even perturbations does not depend on the speed of inhibition,  $\tau$.  However, neutral stability still does depend on $\tau$, as it is possible that other eigenvalues associated with even perturbations may have positive real part. Looking at the other eigenvalues, we have two negative ones, $\lambda_{-} = -\frac{1}{\tau}$, corresponding to perturbations in $\M_1$ and $\N_1$.  Therefore, if we only perturb the inhibitory network, then solutions will be attracted back toward the fixed bump solutions.  The final two eigenvalues can be analyzed by examining
\begin{equation}\label{eq:lambdavstau}
\lambda_{\pm} = \frac{1}{2}\left(1-\frac{1}{\tau}\right) \pm \frac{1}{2\tau}\sqrt{\left(1+\frac{64}{\pi^2}\right)\tau^2 - 6\tau + 1}.
\end{equation}
Note that if $\tau = \frac{\pi^2}{16}$, then we obtain one more zero eigenvalue, however the corresponding eigenvector is the zero vector.  We plot the two eigenvalues determined by Eq.~(\ref{eq:lambdavstau}) in Fig.~\ref{fig:lambdavstau}, showing $\lambda_-$ is negative for all $\tau$ whereas $\lambda_+$ is only negative when $\tau < \frac{\pi^2}{16}$.  Thus we can ensure that all eigenvalues are zero or have negative real part as long as $\tau < \frac{\pi^2}{16}$.  Note that for a certain range of values for $\tau$ we have a non-zero imaginary component in $\lambda$, however both eigenvalues have negative real part so that oscillatory instabilities never arise. 

When $\tau = 0$, then the $u$ equation is equivalent to the single population by letting $w_1 = w_{ee}$.  The analysis we performed on the system (\ref{eq:singlenet}) then applies, and we can derive the same conditions for neutral stability. Thus, we conclude that one of the major differences between networks with one and two populations is that it is possible to destabilize stationary bumps with sufficiently slow inhibition ($\tau$ large enough). In the case of two populations, the restrictions required to derive a network possessing a line attractor of bump amplitudes generate a relationship implying a monotone increasing correspondence between the strength of excitation and inhibition. 

\begin{figure}\label{fig:lambdavstau}
\begin{center}
\includegraphics[scale=1]{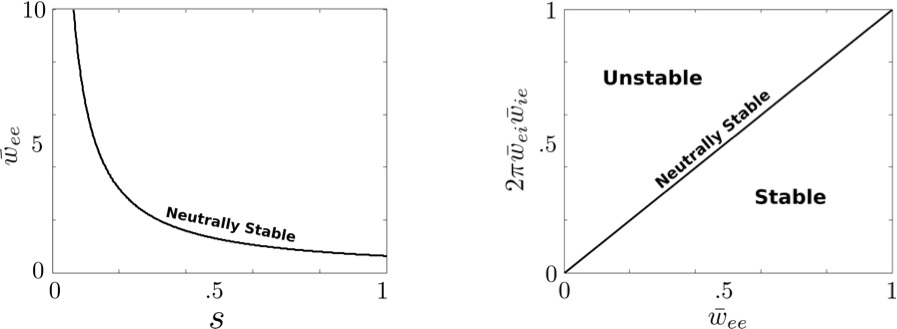}
\end{center}
\caption{Stability regions of stationary bumps in two population network, whose amplitudes are determined by equations (\ref{eq:eff_weight}), (\ref{eq:Acontinuum}). Eigenvalues that determine linear stability are given by (\ref{twopopchar}). (\textbf{A}) The line indicates all the values of $\wee$ and $s$ that induce neutral stability in the network.  (\textbf{B}) The line $\wee = 2\pi\wie\wie$ divides parameter space into unstable region (excess inhibition) and stable region (excess excitation).  We achieve neutral stability when excitatory and inhibitory strengths are balanced properly. }\label{fig:stabilityparam}
\end{figure}

\begin{figure}
\begin{center}
\includegraphics[scale=1]{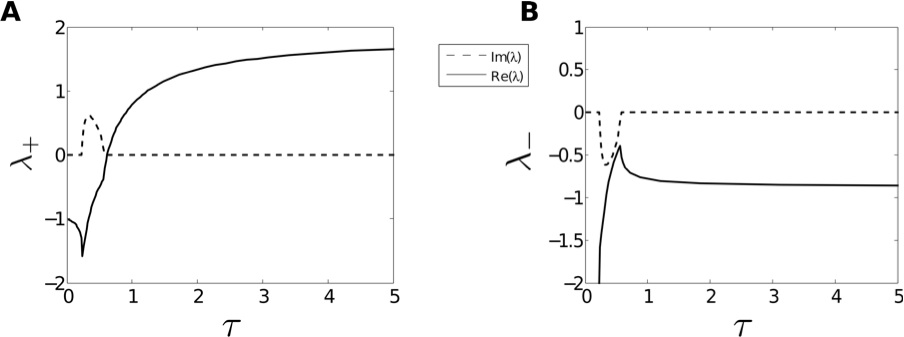}
\end{center}
\caption{Dependence of eigenvalues, defined by equation (\ref{eq:lambdavstau}) plotted against $\tau$ for ({\bf A}) $\lambda_+$ and ({\bf B}) $\lambda_-$.}\label{fig:lambdavstau}
\end{figure}

\subsection{Stochastic motion of bumps in the two population network}

\begin{figure}
\begin{center}
\includegraphics[scale = 1]{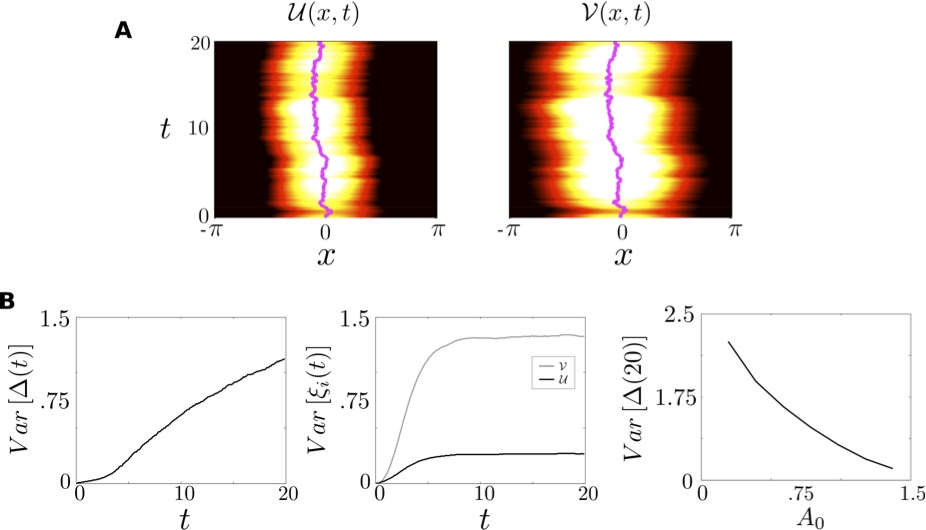}
\end{center}
\caption{Diffusion of bump solution in the two population E-I network.  (\textbf{A}) Space-time plot of bump during the delay period for both the excitatory network, $\mathcal{U}$, and inhibitory network, $\mathcal{V}$.  (\textbf{B}) Variances of the bump center (left) and peak of bump (middle) as well as the variance for $\Delta(t)$, evaluated at $t = 20$, as a function of initial amplitude.  Other parameters used: $\varepsilon = .01$, $s = 2/\pi$, $\wee = \wei = 1$, $\wie = 1/2\pi$, $\tau = .4$, $A_1 = \pi/4$}\label{fig:stochbumpei}
\end{figure}

As in the case of the single population network, we now study how fluctuations affect the motion of bumps in the network (\ref{eq:einet}). To do so, we consider a phenomenological model that incorporates noise as an additive term in a Langevin equation
\begin{align}
\d {\mc U} (x,t) =& \left[ - {\mc U}(x,t) + \int_{- \pi}^{\pi} w_{ee}(x,y) f({\mc U}(y,t)) \d y - \int_{- \pi}^{\pi} w_{ie}(x,y) {\mc V}(y,t) \d y  \right] \d t + \ve^{1/2} \d W(x,t) \nonumber \\
\tau \d {\mc V}(x,t) = & \left[ - {\mc V}(x,t) + \int_{- \pi}^{\pi} w_{ei}(x,y) f({\mc U}(y,t)) \d y \right] \d t    \label{stoch2}
\end{align}
with
\begin{align}
\langle \d W(x,t) \rangle = 0, \ \ \ \ \langle \d W(x,t) \d W(y,s) \rangle = C(x-y) \delta (t-s)  \d t \d s,
\end{align}
where $\ve$ parameterizes the level of noise. Note that, because we could convert the system in Eq.~(\ref{stoch2}) into a single second order stochastic differential equation, it would be redundant to include a noise term in the ${\mc V}$ equation. Rather than performing an asymptotic analysis on the system (\ref{stoch2}), we now briefly present the results of numerical simulations, showing that bumps diffuse in a similar way to the single population network (see Fig. \ref{fig:stochbumpei}A).

\begin{figure}
\begin{center}
\includegraphics[scale=1]{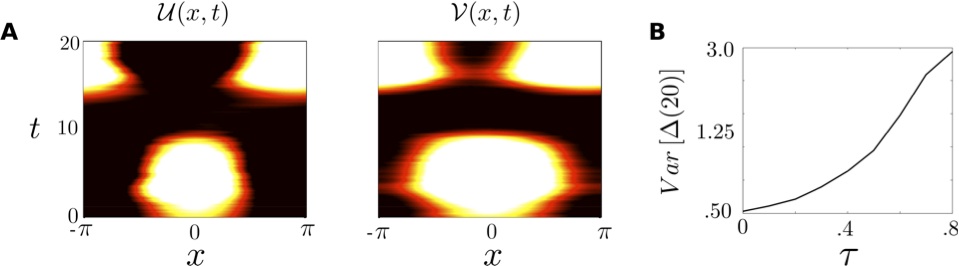}
\end{center}
\caption{(\textbf{A}) Bump extinction occurring when $\tau = .7$.  (\textbf{B}) Variance of $\Delta(t)$ at $t=20$ as a function of $\tau$.  Other values used: $\varepsilon = .01$, $s=2/\pi$, $\wee = \wei = 1$, $\wie = 1/2\pi$, $A_1 = \pi/4$.}\label{fig:bumpextinction}
\end{figure}

First we note that in the limit of fast inhibition, $\tau \to 0$, Eq.~(\ref{stoch2}) is equivalent to Eq.~(\ref{eq:singlestoch}).  Therefore, we would expect identical results for the numerically computed variance of the bump position and amplitude as compared to the single population model in this limit. Moving away from this limit we see slightly different behaviors in the variances of the bump center and maximum as shown in Fig.~\ref{fig:stochbumpei}B. First, we see that the variance of the position $\Delta (t)$ of the bump scales more quickly than in the single population network. One reason for this may be that the excitatory and inhibitory populations are not instantaneously coupled to one another, so the transient behavior of the excitatory population involves more activation than in the single population network. Next, we note that the variance of the bump amplitude saturates more quickly than in the single population network.  Based on the stability analysis we have carried out, we speculate that this saturation may arise because the initial build up in variance is mostly along weakly stable eigendimensions. After this, variance in the amplitude may continue to build up along the remaining neutrally stable eigendimension, happening at a considerably slower pace. However, fully understanding this behavior will require studying the stochastic system in depth, which we leave for future work.  Finally, we still see that the variance of the bump center decreases with increasing initial amplitude as we saw in the single population. Once again, this implies that a signal that initially possesses more certainty will be more robust to dynamic fluctuations during the storage period.

We showed in section \ref{stabilityei} that all the remaining nonzero eigenvalues increase towards zero as $\tau$ increases for $\tau \le \pi^2/16$.  Thus we would expect that as inhibition becomes slower ($\tau$ increases), the variance of the bump's position would increase due to the stability of the bump to certain odd perturbations becomes weaker. We observe this numerically across a broad range of $\tau$ values in Fig. \ref{fig:bumpextinction}B. Essentially, when the inhibitory population does not respond as quickly to transient motion of the excitatory population, noise causes the bump's position to alter more rapidly. Additionally, for $\tau > \pi/16$, one of the eigenvalues associated with the linear stability of the bump becomes positive.  In this case we would expect a total loss of the instantiated bump solution, which we can associated with a total loss of the remembered position, as seen in Fig. \ref{fig:bumpextinction}A with $\tau = .7$. Thus, by considering two separate excitatory and inhibitory populations, we see that the speed of inhibition plays a crucial role in the response of the bump to noise. Previously, \cite{pinto01b} showed that bumps on an unbounded domain are destabilized by oscillatory instabilities for sufficiently slow inhibition. Here, we extend this work by showing non-oscillatory instabilities can occur on the bounded domain of a ring, but these instabilities are still associated with bump extinction.

\section{Discussion}

We have derived conditions under which networks can support bumps with a continuum of amplitudes. While the location of the bump represents the stored location of a stimulus, its amplitude can represent the certainty of this internal representation. These stationary bump solutions are neutrally stable to perturbations in position as well as amplitude. Our analysis shows  that recurrent excitation must be balanced by inhibition to obtain a network that supports such neutrally stable bumps. When considering the effects of noise, we find that the bump diffuses away from its initial position. Not only does the position in orientation space change, but so does the amplitude. Using asymptotic approximations, we can relate the parameters of the model to the effective amount of diffusion the bump will experience. Importantly, bumps with larger initial amplitude diffuse less than bumps with smaller initial amplitudes. Therefore, the amount of certainty initially attached to the stored stimulus determines the fragility of the memory.

We believe this work contributes to the established claims concerning the importance of tuning excitation and inhibition in cortical networks to support flexible computations for cognitive tasks \citep{brunel03,haider06,yizhar11}. Note that by deriving conditions under which a network supports a neutrally stable line attractor, we  tune parameters of the model so its dynamics lie right at a bifurcation, hence moving the system to criticality. Pharmacological manipulations of cortical networks have recently revealed that a precise balance of excitation and inhibition in cortical networks is crucial for criticality, and in this state a network can maximize the range of inputs it can process \citep{shew09}. The fact that balancing excitation and inhibition in a network can lead to an increase in the rate of information transfer was originally shown by the work of \cite{vreeswijk96}. As opposed to the work of \cite{shew09}, which examines the total transient activation of a network in response to stimuli, we are considering the persistent spatially-dependent response of a network encoding cue position. Nonetheless, we note that by balancing excitation and inhibition, we have created networks that can transfer additional information about the certainty of a stimulus. Without this balance, the amplitude of the bump would always relax to a single value. 

The networks we considered are not only useful for studying spatial working memory tasks, but could also be used as neural circuit models of decision making \citep{gold02}. Balancing synaptic feedback with the timescale of synaptic decay is a well-established way to obtain model networks capable of performing two-alternative forced choice (2AFC) tasks \citep{wang02,bogacz06}. External inputs can then be slowly integrated along the direction of their bias. In 2AFC tasks, there are only two possible directions \citep{gold02}. In this work, we have developed a model capable of integrating inputs that can be biased to one of a continuum of dimensions around a ring. Thus, we suggest the two-dimensional attractor we have derived in this work may be ideal for integrating information in the presence of a continuum of alternatives. Recent recordings from lateral interparietal cortex suggest there are neurons whose firing rates climb in correspondence to an animal's certainty about one of a multitude of decisions \citep{churchland08}. Subsequent computational work suggested an entire circuit may then be capable of sequentially updating a probability function that provides the instantaneous likelihood of each alternative being true, based on accumulated input \citep{beck08}. Our model may provide a complementary network implementation of a decision making circuit. Of course, depending on the form of the synaptic weight function, the accumulating persistent activity would depend differently on the timing of inputs. We will explore these issues in future studies.


\bibliography{neutral}
\bibliographystyle{spbasic}      

\end{document}